\documentclass[12pt]{iopart}
\usepackage[T1]{fontenc}
\usepackage[latin9]{inputenc}
\usepackage{bm}
\usepackage{amsbsy}
\usepackage{amssymb}
\usepackage{esint}

\makeatletter
\usepackage{iopams}
\usepackage{setstack}

\makeatother

\begin{document}

\title{Phase-space representations of thermal Bose-Einstein condensates }

\author{King Ng$^{1}$, Rodney Polkinghorne$^{1}$, Bogdan Opanchuk$^{1}$,
Peter D. Drummond$^{1,2}$}

\address{$^{1}$Centre for Quantum and Optical Science, Swinburne University
of Technology, Melbourne 3122, Australia}

\address{$^{2}$Institute of Theoretical Atomic, Molecular and Optical Physics
(ITAMP), Harvard University, Cambridge, Massachusetts, USA. }

\ead{pdrummond@swin.edu.au}
\begin{abstract}
Phase-space methods allow one to go beyond the mean-field approximation
to simulate the quantum dynamics of interacting fields. Here, we obtain
a technique for initializing either Wigner or positive-P phase-space
simulations of Bose-Einstein condensates with quantum states at a
finite temperature. As a means to calculate the initial states, we
introduce the idea of a nonlinear chemical potential, which removes
the zero-momentum phase-noise divergences of Bogoliubov theory to
give a diagonal Hamiltonian. The resulting steady-state quantum theory
is then directly applicable to the calculations of initial conditions
for quantum simulations of BEC dynamics using phase-space techniques.
These methods allow efficient and scalable simulation of large Bose-Einstein
condensates. We suggest that nonlinear chemical potentials may have
a general applicability to cases of broken symmetry. 
\end{abstract}

\submitto{\jpa}

\maketitle

\section{Introduction}

Since his original method for treating superfluidity was first published,
Bogoliubov's approach \cite{Bogoliubov1947} has had many useful applications
to both superfluidity and superconductivity. One problem with this
method is that it either diverges for zero-momentum phase fluctuations,
or does not fully diagonalize the Hamiltonian. These divergences are
inherent in the idea of symmetry breaking. Because this technique
is central to condensed matter physics and QCD, a variety of nonlinear
operator transformations have also been introduced to overcome the
difficulties of the Bogoliubov approach \cite{girardeau1959theory,MorganGapless2000}.
In this paper, we remove the zero-momentum divergence in a simpler
way, to obtain a unitary transformation that diagonalizes the grand
canonical Hamiltonian. As an important application, we show how this
technique can be used to obtain stochastic initial conditions for
quantum phase space simulations of Bose-Einstein condensate (BEC)
dynamics, including correlations and entanglement.

The physical reason for the divergence is clear. Bogoliubov's approach
assumes that the phase symmetry of the Hamiltonian is broken in thermal
equilibrium. This cannot be rigorously correct. The BEC Hamiltonian
commutes with the number operator. Hence, energy eigenstates must
have a well-defined particle number. The Noether theorem \cite{noether1971invariant}
shows the relationship between number conservation and symmetry: a
quantum state with broken phase symmetry can be neither a number eigenstate
nor an energy eigenstate. We overcome this through an approach of
phase-averaging over the broken symmetry phases.

Phase-averaging by itself would not create a diagonal Hamiltonian
transformation. We achieve this by combining known results on quantum
phase-diffusion \cite{blaizot1986quantum,lewenstein1996quantum},
with a nonlinear chemical potential term. This is similar to the usual
linear chemical potential term, and neither type of chemical potential
alters the exact quantum dynamical equations. However, a nonlinear
chemical potential alters \emph{the initial grand canonical ensemble},
which does change the effect of linearization in a subtle way. The
idea is not unreasonable, since the energy of particle exchange with
a reservoir is nonlinear in the particle number. A related approach
of using a second Lagrange multiplier is known \cite{yukalov2006gapless}.

Such issues were not important in early work on superfluidity with
large particle numbers, but they are significant in mesoscopic experiments,
where number fluctuations, dynamics and entanglement are experimentally
measurable. Mesoscopic Bose condensates are now produced with ultra-cold
atomic gases \cite{AndersonBEC1995,Davis1995BEC}. Quantum dynamics
and thermodynamics is being investigated on mesoscopic scales with
these techniques \cite{AndrewsInterference1997,GreinerCollapse2002,GrundControl2009,Kitagawa2010Dynamics,MazetsThermal2010,kitagawa2011dynamics,BarmettlerCorrelations2012,Cheneau2012,Gring2012Prethermalization,HungCosmology2013,langen2013local,deng2018superadiabatic}.
This means that the existence or otherwise of symmetry breaking is
relevant and observable. Yet Bogoliubov's approach still has considerable
utility.

Related approaches to this issue include using modified field operators
\cite{gardiner1997particle,castin1998low}, or number conserving approximations
to the initial quantum state \cite{Leggett2001,jiang2016particle}.
These methods are not so readily applicable to multimode quantum dynamics
owing to its exponential complexity. Here we show that physically
correct results may be obtained in simpler ways. Retaining the Bogoliubov
transformation allows the use of operator mappings into phase-space,
which simplifies the calculation of dynamical effects. The standard
commutation properties and phase-space mappings of bosonic field operators
are retained with this approach.

The reason why this is important is that many current experiments
in BEC are both highly dynamical and quantum limited. Such questions
as the use of a BEC as a quantum field theory analog to quantum tunneling
in the early universe \cite{Opanchuk2013adp-525-866,Fialko2017},
the limits to matter-wave interferometry \cite{egorov2011long}, and
quantum field entanglement \cite{fadel2018spatial,kunkel2018spatially}
have all become the subjects of experimental investigation. For these
problems, neither mean field theory nor perturbation theory is directly
useful. The Gross-Pitaevskii equation \cite{Gross1961,pitaevskii1961vortex}
ignores quantum fluctuations, and cannot be used to predict spin squeezing
or entanglement \cite{kunkel2018spatially,fadel2018spatial}. There
are more sophisticated mean field methods, but these have the problem
that it is difficult to satisfy the constraints of obeying conservation
laws, while obtaining a gapless excitation spectrum \cite{griffin1996conserving}.

Instead, it is often most practical to use phase space dynamical quantum
simulations. However, the calculation of an appropriate thermal initial
quantum state is essential. The present approach gives a direct estimate
of the initial quantum state in an equilibrium BEC with a technique
that removes divergences, which allows one to represent the initial
thermal state using a phase-space method. This can then be used for
calculating quantum dynamics. The results obtained here are for low
temperature BECs with repulsive interactions. Modifications would
be required near the critical temperature \cite{andersen2004theory},
or for attractive interactions.

\section{Hamiltonian and chemical potential}

We assume a standard Bose gas Hamiltonian $\hat{H}$ \cite{1971-Fetter-Quantum},
with an external potential $U(x)$, an S-wave coupling $g$ and mass
$m$: 
\begin{equation}
\hat{H}=\int d\bm{x}\left[\hat{\Psi}^{\dagger}\left(-\frac{\hbar^{2}\nabla^{2}}{2m}+U(\bm{x})\right)\hat{\Psi}+\frac{g}{2}\hat{\Psi}^{\dagger2}\hat{\Psi}^{2}\right]\,.
\end{equation}

Here $\hat{\Psi}=\hat{\Psi}\left(\bm{x}\right)$ is a bosonic field,
$\bm{x}$ is a coordinate in one or more dimensions, we define $d\bm{x}\equiv dx_{1}dx_{2}\ldots$.,
and the integration is over a finite volume V. For simplicity, we
ignore internal spin, although this can be included without changing
our main arguments. We suppose there is a momentum cutoff, otherwise
renormalization counter-terms are required. We will suppress field
arguments $(\bm{x})$ where there is no ambiguity. A standard mode
expansion can be used to diagonalize the noninteracting part of the
Hamiltonian, which has the form: 
\begin{equation}
\hat{\Psi}\left(\bm{x}\right)=\sum_{\bm{k}}u_{\bm{k}}^{(0)}\left(\bm{x}\right)\hat{a}_{\bm{k}}\,,\label{eq:standard-mode}
\end{equation}
where $\hat{a}_{\bm{k}}$ is a bosonic annihilation operator with
the usual Bose commutation relations, and $u_{\bm{k}}^{(0)}\left(\bm{x}\right)$
is an orthonormal set of free-field mode functions.

To obtain the initial number conserving quantum density matrix $\hat{\rho}$,
we will assume that it has the form of a phase-average over a condensate
phase $\phi$, such that 
\begin{equation}
\hat{\rho}=\frac{1}{2\pi}\int d\phi\hat{\rho}\left(\phi\right)\,.
\end{equation}
We will also introduce a condensate wave-function $\Psi_{\phi}$ that
is conditioned on the phase $\phi$, where $\Psi_{\phi}=\Psi_{0}\left(\bm{x}\right)\exp\left(i\phi\right)$.
Here $\Psi_{0}\left(\bm{x}\right)$ will be evaluated later. Operationally,
one may regard this as corresponding to a procedure where the phase
is measured in a way that projects out $\hat{\rho}\left(\phi\right)$.
Yet without an external injected signal, the absolute phase is not
accessible, and so we will not be concerned with measuring the absolute
phase.

Similar techniques are used to model the quantum state of a laser.
As an example often used in laser physics, one might assume that $\hat{\rho}\left(\phi\right)=\left|\Psi_{\phi}\right\rangle \left\langle \Psi_{\phi}\right|$
where $\left|\Psi_{\phi}\right\rangle $ is a Glauber coherent state
\cite{Glauber1963,Bargmann1961}. This example is exactly equivalent
to a Poissonian distribution of number states, showing that phase
and number distributions are complementary. While the BEC case is
more complex, related approaches have been used in variational calculations
\cite{navez1998macroscopic}. At the same time, distributions over
a range of particle numbers are used in statistical mechanics in the
form of the grand canonical ensemble, and they are universally found
in current BEC experiments \cite{chuu2005direct}. These are more
generally appropriate distributions to use than number states, which
are experimentally difficult to access at large particle number.

Since number is conserved, any initial thermal state is typically
a mixture of number eigenstates, and experiments are carried out by
changing a number conserving part of the Hamiltonian. Because it commutes
with the Hamiltonian, one can add or subtract from the Hamiltonian
any function of the total number $\hat{N}$, without altering the
evolution of any number eigenstate. This is called a chemical potential
term. Such terms cause a relative energy change and hence a phase
shift between states of different total number, but the absolute phase
is not observable because of super-selection rules. It is common in
Bogoliubov theory to use this freedom to include a linear chemical
potential. Here we extend this procedure to include a nonlinear chemical
potential.

We note that an unobservable phase only occurs when the Hamiltonian
describes the entire Bose condensate, and not just part of it. By
contrast, a split condensate \emph{does} have a well-defined relative
phase between the portions it is split into. Our method can describe
such cases. However, the Hamiltonian must describe the entire system,
including both of the split condensates. In representing these or
other experiments, $N=\left\langle \hat{N}\right\rangle $ is usually
known. Higher moments of the number may be measured as well \cite{chuu2005direct},
and these will constrain the initial ensemble.

To obtain the initial state, we will employ a type of grand canonical
Hamiltonian $\hat{K}$ that includes a nonlinear chemical potential
term, $\hat{K}_{\bm{\mu}}$, so $\hat{K}=\hat{H}+\hat{K}_{\bm{\mu}}$.
The nonlinear chemical potential introduced here is a Taylor expansion
of a generalized chemical potential $\mu(\hat{N})$, and is defined
so that, to second order: 
\begin{equation}
\hat{K}=\hat{H}-\mu_{1}\hat{N}-\frac{\mu_{2}}{2}\hat{N}^{2}\,.
\end{equation}

This modification does not change any observable dynamical property.
In fact it is conventional to use a linear chemical potential in the
thermodynamic limit of large particle number $N$. Yet reservoirs
generally are nonlinear, and they can have a significant effect in
mesoscopic thermodynamics. Hence, we will use $\hat{K}$ rather than
$\hat{H}$ in our calculations of the statistical ensemble. The argument
holds even if there are several spin components. We will show that
this alteration leads to an improved understanding of the Bogoliubov
linearization method.

Our motivation for choosing this form of $\hat{K}$ is to obtain a
diagonal Bogoliubov Hamiltonian, including the zero momentum mode.
However, there is a fundamental issue as well. A chemical potential
is intended to model the energy changes when a particle is transferred
to or from a reservoir. The usual choice of a linear term is not always
appropriate for a nonlinear interactions in a BEC. As we will show,
the present choice of a nonlinear chemical potential results in an
effective Hamiltonian that is stable for small fluctuations. There
is no energy gap in the limit of long wavelength condensate fluctuations,
which is generally regarded as essential to a theoretical description
of a Bose condensate \cite{hugenholtz1959ground}.

With this added term, we first need to understand if the grand canonical
Hamiltonian is bounded below in the case of a repulsive inter-particle
potential, or positive scattering length. This is because we wish
to use $\hat{K}$ to calculate initial states in thermal equilibrium.
In a homogeneous BEC, we show below that the linearized phase divergence
is suppressed by choosing the chemical potential term to be completely
nonlinear, with $\mu_{2}=g/V$. As a result, the nonlinear chemical
potential counter-term cancels the many-body energy shift of the nonlinear
interaction term for the case of a uniform coherent state, where $g^{(2)}(0)=1$,
apart from terms of $O(1/N)$ for $N$ particles. The conventional
linear chemical potential only cancels the single-particle energy
shift.

An obvious question is: why add a conserved quantity to the Hamiltonian?
Similar questions arise with gauge transformations. The answer is
that it is often more convenient to modify a Hamiltonian before using
approximations. This is mathematically rigorous as long as the exact
dynamics is invariant. In this case, the present approach can be used
to diagonalize the linearized Hamiltonian. We do not have to drop
terms from the mode expansion as with previous approaches, although
it should be emphasized that the main advantage is in calculating
initial states, since the nonlinear dynamics is unchanged if the full
Hamiltonian is used for this.

In the following sections, we will prove that the resulting grand
canonical Hamiltonian is stable for small fluctuations. This is sufficient
to obtain an approximate theory for calculating the initial state.
Linearization is not required for our dynamical calculations. Quantitative
results of these are presented elsewhere. These calculations are carried
out with the nonlinear Hamiltonian using phase-space techniques. For
dynamical results, no chemical potential is needed and the question
of boundedness is irrelevant. 

\section{Small fluctuation expansion}

Following the methods of Bogoliubov \cite{Bogoliubov1947}, we assume
that in the phase-dependent initial density matrix component $\hat{\rho}\left(\phi\right)$,
there are small quantum field fluctuations $\delta\hat{\Psi}$ around
a condensate field $\Psi_{\phi}\left(\bm{x}\right)$, such that 
\begin{equation}
\hat{\Psi}=\Psi_{\phi}+\delta\hat{\Psi}\,.
\end{equation}
This is equivalent to a unitary transformation, since the unitary
displacement operator 
\begin{equation}
\hat{D}\left[\Psi_{\phi}\right]=\exp\left[i\int d\bm{x}\left(\hat{\Psi}^{\dagger}\Psi_{\phi}-\hat{\Psi}\Psi_{\phi}^{*}\right)\right]
\end{equation}
has the well known effect \cite{Glauber1963} that: 
\begin{equation}
\delta\hat{\Psi}=\hat{D}\hat{\Psi}\hat{D}^{\dagger}=\hat{\Psi}-\Psi_{\phi}.
\end{equation}

One may interpret $\phi$ as the phase obtained from spontaneous symmetry-breaking.
More precisely, $\phi$ is the phase of the conditional density matrix
component $\hat{\rho}\left(\phi\right)$ that we wish to analyze.
We do \emph{not} assume that $\Psi_{\phi}=\left\langle \hat{\Psi}\right\rangle $.
This would lead to a contradiction.

In the case of a thermal ensemble, we must generate a distribution
of number eigenstates for the overall density matrix, $\hat{\rho}$,
which cannot have a well-defined phase. To obtain any observable result,
one must average over the phase $\phi$, using the distribution $\hat{\ensuremath{\rho}}\left(\phi\right)$,
so that $\left\langle \hat{\Psi}\right\rangle =0$ \textendash{} as
required by quantum mechanics. From now on, we will calculate with
just one of these possible phases $\phi$, and it is no limitation
to choose $\phi=0$. Since the final results of number-conserving
observables do not depend on $\phi$, one choice is as good as another,
and explicit phase-averaging of the observables is not required.

We note that the condensate field $\Psi_{\phi}$ can be expanded in
modes, just as in Eq (\ref{eq:standard-mode}), so that:

\begin{equation}
\Psi_{\phi}\left(\bm{x}\right)=\sum_{\bm{k}}u_{\bm{k}}^{(0)}\left(\bm{x}\right)\alpha_{\bm{k}}^{(0)}\,.\label{eq:standard-mode-1}
\end{equation}

\subsection{Hamiltonian expansion}

On introducing $n_{0}=\left|\Psi_{0}\right|^{2}$, we define the condensate
number as: 
\begin{equation}
N_{0}=\int d\bm{x}n_{0}\left(\bm{x}\right)\,.
\end{equation}
The resulting number density operator, $\hat{n}\left(\bm{x}\right)=\hat{\Psi}^{\dagger}\hat{\Psi}$,
is 
\begin{equation}
\hat{n}\left(\bm{x}\right)=n_{0}\left(\bm{x}\right)+\Psi_{0}^{*}\delta\hat{\Psi}+\Psi_{0}\delta\hat{\Psi}^{\dagger}+\delta\hat{\Psi}^{\dagger}\delta\hat{\Psi}\,.
\end{equation}
It is convenient to introduce a Bose quadrature operator as the zero
momentum part of $\delta\hat{\Psi}$. Defining a normalized mode function,
$\psi_{0}=\Psi_{0}/\sqrt{N_{0}}$ we introduce the condensate quadrature
operator as:
\begin{equation}
\hat{P}\equiv\int d\bm{x}\left(\psi_{0}^{*}\delta\hat{\Psi}+\psi_{0}\delta\hat{\Psi}^{\dagger}\right)/\sqrt{2}\,.\label{eq:quadrature}
\end{equation}
The number operator can now be rewritten to second order in the quantum
field fluctuations, giving:
\begin{equation}
\hat{N}=N_{0}+\hat{P}\sqrt{2N_{0}}+\int d\bm{x}\delta\hat{\Psi}^{\dagger}\delta\hat{\Psi}\,.\label{eq:NumberOperator}
\end{equation}
Similarly, the Hamiltonian can be expanded order by order in $\delta\hat{\Psi}$,
so that 
\begin{equation}
\hat{H}=\sum_{n=0}^{4}\hat{H}^{(n)}\,.
\end{equation}

The lowest order terms in $\hat{H}$ are then:

\begin{eqnarray}
\hat{H}^{(0)} & = & \int d\bm{x}\Psi_{\phi}^{*}\left(-\frac{\hbar^{2}\nabla^{2}}{2m}+U\left(\bm{x}\right)+\frac{g}{2}n_{0}\right)\Psi_{\phi},\\
\hat{H}^{(1)} & = & \int d\bm{x}\left(\delta\hat{\Psi}\mathcal{H}\Psi_{0}^{*}+\delta\hat{\Psi}^{\dagger}\mathcal{H}\Psi_{0}\right),\nonumber \\
\hat{H}^{(2)} & = & \hat{H}_{0}+\frac{g}{2}\int d\bm{x}\left(\Psi_{0}^{*}\delta\hat{\Psi}+\Psi_{0}\delta\hat{\Psi}^{\dagger}\right)^{2}.\nonumber 
\end{eqnarray}

Here the differential operator $\mathcal{H}$ is the effective single-particle
Hamiltonian 
\begin{equation}
\mathcal{H}=-\frac{\hbar^{2}\nabla^{2}}{2m}+U\left(\bm{x}\right)+gn_{0}\left(\bm{x}\right)\,,\label{eq:single-particleH}
\end{equation}
and the mean-field quantum Hamiltonian $\hat{H}_{0}$ is 
\begin{equation}
\hat{H}_{0}=\int d\bm{\bm{x}}\delta\hat{\Psi}^{\dagger}\mathcal{H}\delta\hat{\Psi}\,.
\end{equation}

We consistently omit small c-number corrections due to commutation
relations, as they have no effect either on the dynamics or on the
density matrix. To include the chemical potential term to second order,
let:
\begin{equation}
\hat{K}^{(n)}=\hat{H}^{(n)}+\hat{K}_{\bm{\mu}}^{(n)}.
\end{equation}
If we define an effective linear chemical potential as $\mu_{e}=\mu_{1}+\mu_{2}N_{0}$,
one finds that to second order, the additional terms are: 
\begin{eqnarray}
\hat{K}_{\bm{\mu}}^{(0)} & = & -\mu_{1}N_{0}-\mu_{2}N_{0}^{2}/2\,,\nonumber \\
\hat{K}_{\bm{\mu}}^{(1)} & = & -\mu_{e}\hat{P}\sqrt{2N_{0}}\,,\nonumber \\
K_{\bm{\mu}}^{(2)} & = & -\mu_{2}\hat{P}^{2}N_{0}-\mu_{e}\int d\bm{x}\delta\hat{\Psi}^{\dagger}\delta\hat{\Psi}\,.
\end{eqnarray}

\subsection{Canceling linear terms }

The total linear term including the chemical potential contributions
can be written as: 
\begin{equation}
\hat{K}^{(1)}=\int d\bm{x}\left(\delta\hat{\Psi}^{\dagger}\left[\mathcal{H}-\mu_{e}\right]\Psi_{0}+h.c.\right)\,.
\end{equation}
If we choose $\mu_{e}$ and $\Psi_{0}$ such that the $\mathcal{H}\Psi_{0}=\mu_{e}\Psi_{0}$,
then this linear term is canceled completely. Thus, with the correct
choice of chemical potential, one can remove the linear term in the
Hamiltonian, leaving only quadratic terms after linearization. The
second-order term in this expansion of the grand canonical Hamiltonian
is 
\begin{eqnarray}
\hat{K}^{(2)} & = & \hat{H}_{0}-\mu_{2}\hat{P}^{2}N_{0}\\
 &  & +\frac{g}{2}\int d\bm{x}n_{0}\left(\bm{x}\right)\left(\Psi_{0}^{*}\delta\hat{\Psi}+\Psi_{0}\delta\hat{\Psi}^{\dagger}\right)^{2}.\nonumber 
\end{eqnarray}

This is identical to results found in the literature for the case
of either a homogeneous or trapped Bose gas\cite{Griffin1997,Fetter1972,DoddCollective1998,Bogoliubov1947,you1997low,singh1996collective,Mora2003quasicondensates,lewenstein1996quantum,leggett2001bose},
except for the additional nonlinear chemical potential term proportional
to $\mu_{2}.$

\section{Bogoliubov transformation}

Next, we make a canonical transformation, following standard techniques.
This \cite{Bogoliubov1947,Fetter1972,lewenstein1996quantum,mine2005relation}
is a modified version of the standard expansion of Eq (\ref{eq:standard-mode}),
defined as 
\begin{equation}
\hat{b}_{\bm{k}}=\int d\bm{x}\left[u_{\bm{k}}^{*}\left(\bm{x}\right)\delta\hat{\Psi}\left(\bm{x}\right)+v_{\bm{k}}\left(\bm{x}\right)\delta\hat{\Psi}^{\dagger}\left(\bm{x}\right)\right]\,,\label{eq:Bogoliubov}
\end{equation}
where $\hat{b}_{\bm{k}}$ is a bosonic mode operator, so that $\left[\hat{b}_{\bm{k}},\hat{b}_{\bm{q}}\right]=0$,
$\left[\hat{b}_{\bm{k}},\hat{b}_{\bm{q}}^{\dagger}\right]=\delta_{\bm{k},\bm{q}}$,
and in general $u,v$ depend on the condensate field. Here $\bm{k}$
is one of a set of mode indices, which are not generally the momenta,
but may include angular momenta or other eigenvalues in spherical
traps.

This is another canonical transformation \cite{Yuen1976,schumaker1985new,drummond2013quantum},
since if $\delta\hat{a}$ is the shifted mode operator, and $\hat{U}_{S}\left(\bm{u},\bm{v}\right)$
is the squeezing operator equivalent to Eq (\ref{eq:Bogoliubov}),
then: 
\begin{equation}
\hat{b}_{\bm{k}}=\hat{U}_{S}\delta a_{\bm{k}}\hat{U}_{S}^{\dagger}.
\end{equation}
The mode functions $u_{\bm{k}}$ and $v_{\bm{k}}$ are derived below,
and the purpose of the transformation is to diagonalize the quadratic
approximation to the Bose gas Hamiltonian. To maintain commutation
relations \cite{Fetter1972}, one must require: 
\begin{equation}
\sum_{\bm{k}}\left[u_{\bm{k}}\left(\bm{x}\right)u_{\bm{k}}^{*}\left(\bm{x}'\right)-v_{\bm{k}}^{*}\left(\bm{x}\right)v_{\bm{k}}\left(\bm{x}'\right)\right]=\delta(\bm{x}-\bm{x}')
\end{equation}

and:
\begin{equation}
\sum_{\bm{k}}\left[u_{\bm{k}}\left(\bm{x}\right)v_{\bm{k}}^{*}\left(\bm{x}'\right)-v_{\bm{k}}^{*}\left(\bm{x}\right)u_{\bm{k}}\left(\bm{x}'\right)\right]=0\,.
\end{equation}
The inverse is: 
\begin{equation}
\delta\hat{\Psi}\left(\bm{x}\right)=\sum_{\bm{k}}\left[\hat{b}_{\bm{k}}u_{\bm{k}}\left(\bm{x}\right)-\hat{b}_{\bm{k}}^{\dagger}v_{\bm{k}}^{*}\left(\bm{x}\right)\right].
\end{equation}

If we wish to express this in terms of the original free-field modes
of Eq (\ref{eq:standard-mode}), then clearly one has: 
\begin{equation}
\hat{a}_{\bm{k}}=u_{\bm{k}\bm{q}}\hat{b}_{\bm{q}}-v_{\bm{k}\bm{q}}^{*}\hat{b}_{\bm{k}}^{\dagger}\,,
\end{equation}
where: 
\begin{eqnarray}
u_{\bm{k}\bm{q}} & = & \int d\bm{x}u_{\bm{k}}^{(0)*}\left(\bm{x}\right)u_{\bm{q}}\left(\bm{x}\right),\nonumber \\
v_{\bm{k}\bm{q}} & = & \int d\bm{x}u_{\bm{k}}^{(0)}\left(\bm{x}\right)v_{\bm{q}}\left(\bm{x}\right)\,.
\end{eqnarray}

In the context of quantum optics, this transformation is called a
squeezing transformation, and when applied to the vacuum state it
generates a squeezed state \cite{walls1983squeezed,drummond2013quantum}.
We note the following normalization conditions: 
\begin{equation}
\int d\bm{x}\left[u_{\bm{k}}^{*}\left(\bm{x}\right)u_{\bm{q}}\left(\bm{x}\right)-v_{\bm{k}}^{*}\left(\bm{x}\right)v_{\bm{q}}\left(\bm{x}\right)\right]=\delta_{\bm{k},\bm{q}},
\end{equation}
and:
\begin{equation}
\int d\bm{x}\left[u_{\bm{k}}\left(\bm{x}\right)v_{\bm{q}}\left(\bm{x}\right)-v_{\bm{k}}\left(\bm{x}\right)u_{\bm{q}}\left(\bm{x}\right)\right]=0\,.
\end{equation}

Early work on this problem omitted the $\bm{k}=0$ term \cite{Fetter1972},
which we define as the label of the lowest energy or condensate mode
term. There was a reason for this. A diagonal expansion with $\bm{k}=0$
included is not consistent with the usual Bose gas Hamiltonian $\hat{H}$,
even with a linear chemical potential added. In addition, the original
work on this problem was concerned with collective modes which have
$\bm{k}\neq0$. It was generally argued that quantum fluctuations
of the condensate mode with $\bm{k}=0$ would become negligible in
the thermodynamic limit of $N\rightarrow\infty$.

This approximation of omittig the $\bm{k}=0$ term is not without
physical significance. In the standard model, with more interacting
fields, the $\bm{k}=0$ or Goldstone mode can become a massive Higgs
boson \cite{higgs1964broken}. Although its mass is not known from
first principles, this is now regarded as experimentally confirmed
\cite{chatrchyan2012observation}. Related phase diffusion effects
in a photonic Bose gas cause soliton quantum squeezing \cite{carter1987squeezing,haus1990quantum,drummond1993quantum}.
This is a quantitative, first-principles prediction, which is verified
in experiments \cite{rosenbluh1991squeezed,corney2006many}. One may
expect similar issues in superconducting nanostructures and Majorana
excitations, which also have broken symmetry \cite{plugge2017majorana},
as well as in many other related problems with broken symmetries. 

In summary, the problem with omitting the $\bm{k}=0$ term is that
when the condensate term is \emph{omitted,} the transformation is
not consistent with the original quantum field commutation relations
\cite{lewenstein1996quantum,mine2005relation}. The physical issue
is that the system has a phase symmetry with an associated Nambu-Goldstone
mode \cite{goldstone1961field,nambu1961dynamical}. This must have
a corresponding quantum operator. While the resulting effects are
typically of order $1/N$, they are not always negligible in current
experiments. Photonic experiments with quantum solitons confirm squeezing
induced by quantum phase diffusion \cite{corney2006many}, with $N\approx10^{7}$
. BEC experiments have lower particle numbers than this, typically
in the range of $10^{2}-10^{6}$, and as a result these effects are
even larger.

Accordingly, one must include the $\bm{k}=0$ mode operator. This
is essential, in order to have a consistent quantum transformation
that accurately treats finite condensates. The issue has been treated
previously \cite{lewenstein1996quantum,mine2005relation}. In the
next subsection we summarize this earlier work. We also show how the
use of a nonlinear chemical potential can solve the problems caused
by including this mode in a BEC calculation. This allows one to include
the $\bm{k}=0$ mode operator in a diagonal linearized Hamiltonian,
for calculations of the initial density matrix.

\subsection{Expansion coefficients}

We will first discuss the general approach, then give an example for
the homogeneous case. The Bogoliubov expansion is a bi-orthogonal
expansion, \cite{Fetter1972,lewenstein1996quantum,mine2005relation}.
This is intended to diagonalize the Hamiltonian. To achieve this goal,
we introduce a modified differential operator, $\mathcal{H}_{e}=\mathcal{H}+gn_{0}\left(\bm{x}\right)-\mu_{e}$.
The coefficients $u_{\bm{k}}$ and $v_{\bm{k}}$ are defined to satisfy
the relation that, for $\bm{k}\neq0$: 
\begin{equation}
\left[\begin{array}{cc}
\mathcal{H}_{e} & -gn_{0}\left(\bm{x}\right)\\
gn_{0}\left(\bm{x}\right) & -\mathcal{H}_{e}
\end{array}\right]\left[\begin{array}{c}
u_{\bm{k}}\left(\bm{x}\right)\\
v_{\bm{k}}\left(\bm{x}\right)
\end{array}\right]=\epsilon_{\bm{k}}\left[\begin{array}{c}
u_{\bm{k}}\left(\bm{x}\right)\\
v_{\bm{k}}\left(\bm{x}\right)
\end{array}\right].\label{eq:BdG}
\end{equation}

The $\bm{k}=0$ or condensate mode functions in the expansion of Eq(
\ref{eq:Bogoliubov}) are given by $u_{0}\left(\bm{x}\right)=\left[\psi_{0}+\Phi_{0}\right]/2$
and $v_{0}\left(\bm{x}\right)=\left[\psi_{0}-\Phi_{0}\right]/2$,
where $\psi_{0},\Phi_{0}$ are mode functions introduced previously
\cite{you1997low,villain1997quantum} to solve the equations:
\begin{eqnarray}
\left(\mathcal{H}_{e}+gn_{0}\right)\Phi_{0} & = & 2\alpha\psi_{0}\nonumber \\
\left(\mathcal{H}_{e}-gn_{0}\right)\psi_{0} & = & 0\,.\label{eq:zero-mode}
\end{eqnarray}
 Here, as in previous sections, $\psi_{0}$ is the normalized condensate
wave-function. The normalization condition that defines $\Phi_{0}$
(and hence $\alpha$) is 
\begin{equation}
\int d\bm{x}\left[\psi_{0}\Phi_{0}^{*}+\psi_{0}^{*}\Phi_{0}\right]=2.
\end{equation}

In special cases, like plane waves, analytic solutions for these two
wave-functions can be found. Otherwise, approximate variational \cite{hu2004analytical}
or numerical solutions \cite{javanainen1996noncondensate,hutchinson1997finite}
are necessary. The corresponding zero momentum or condensate mode
operator is: 
\begin{equation}
\hat{b}_{0}=\left(\hat{P}-i\hat{Q}\right)/\sqrt{2}\,,
\end{equation}
where $\hat{P}$ is the quadrature operator defined in Eq (\ref{eq:quadrature}),
and $\hat{Q}$ is a complementary operator defined such that $\left[\hat{Q},\hat{P}\right]=i$,
and hence: 
\begin{equation}
\hat{Q}=i\int d\bm{x}\left[\Phi_{0}^{*}\delta\hat{\Psi}-\Phi_{0}\delta\hat{\Psi}^{\dagger}\right]/\sqrt{2}\,.
\end{equation}

This gives more than just a diagonalization. There is also an off-diagonal
energy term of $\alpha\hbar\hat{P}^{2}$. The quadratic expansion
for the grand canonical Hamiltonian, including the nonlinear chemical
potential, is therefore 
\begin{equation}
\hat{K}_{0}^{(2)}=\left(\alpha-\mu_{2}N_{0}\right)\hat{P}^{2}+\sum_{\bm{k}}\epsilon_{\bm{k}}\hat{b}_{\bm{k}}^{\dagger}\hat{b}_{\bm{k}}\,.
\end{equation}

We see that a nonlinear chemical potential with 
\begin{equation}
\mu_{2}=\alpha/N_{0}\label{eq:nonlinearChemPotSoln}
\end{equation}
 eliminates the $\hat{P}^{2}$ term, allowing one to obtain a rigorous
diagonalization of the Hamiltonian. This is a consistent, unitary
transformation of the quantum field, with \emph{no} omitted modes.

\subsection{Transformed Hamiltonian and thermal equilibrium state}

Dropping the constant terms, which have no effect on dynamics, and
setting $\epsilon_{0}=0$, the final quadratic grand canonical Hamiltonian
is diagonal:
\begin{equation}
\hat{K}^{(2)}=\sum_{\bm{k}}\epsilon_{\bm{k}}\hat{b}_{\bm{k}}^{\dagger}\hat{b}_{\bm{k}}\,.
\end{equation}

The number operator can be given in terms of the fluctuation terms
in Eq (\ref{eq:NumberOperator}). Substituting $\delta\hat{\Psi}=\sum\left(u_{\bm{k}}\left(\bm{x}\right)\hat{b}_{\bm{k}}-v_{\bm{k}}^{*}\left(\bm{x}\right)\hat{b}_{\bm{k}}^{\dagger}\right)$,
and taking expectation values, one can obtain the total boson number.
The actual number depends on the assumed initial quantum state, which
is not specified yet. 

We suppose that the $\bm{k}=0$ operator $\hat{b}_{0}$ is \emph{not}
coupled to thermal energy reservoirs. It mostly causes number fluctuations,
and therefore is constrained through number conservation laws rather
than through energy conservation. This leads to the assumption that
initially,

\begin{equation}
\hat{\rho}\left(\phi_{0}\right)=\hat{\rho}_{th}\hat{\rho}_{0},
\end{equation}
where $\hat{\rho}_{th}$ is a thermal state for the $k\neq0$ modes.
If one has thermally excited phonons for the other modes, as is common,
then for $\bm{k}\neq0$ 
\begin{equation}
\left\langle \hat{n}_{\bm{k}}\right\rangle \equiv\tilde{n}_{\bm{k}}=\left[e^{\beta\epsilon_{\bm{k}}}-1\right]^{-1}\,.
\end{equation}
This expression is restricted to the $\bm{k}\neq0$ terms, and it
is only approximately valid, since it relies on a linearization approach. 

Here, $\hat{\rho}_{0}$ is the state of the $k=0$ mode. It has a
time evolution under $\hat{K}$ with a zero energy excitation up to
the order of the approximate linearized expansion. This means that
all quantum states $\hat{\rho}_{0}$ are equally stationary, whether
ground-states, thermal states, number states or any other candidate
states. 

One may assume in particular that $\left\langle \hat{P}\right\rangle =0$
for the state of the $k=0$ mode, so that the only number fluctuations
are from quadratic terms. For the case of an atomic BEC, the preparation
uses non-equilibrium evaporative or laser cooling. Hence the true
quantum state is not guaranteed to correspond exactly to either a
traditional canonical or a grand canonical ensemble \cite{Drummond1999}.
In fact, the actual initial quantum state in a BEC may be quite complex,
but the present approach will be appropriate in many cases.

Defining an extended row vector $\hat{\underline{a}}=\left[\bm{a},\bm{a}^{\dagger}\right]$,
with hermitian conjugate $\hat{\underline{a}}^{\dagger}=\left[\bm{a}^{\dagger},\bm{a}\right]^{T}$,
one can obtain an initial symmetrically ordered $2M\times2M$ modal
correlation matrix which can be calculated from the Bogoliubov results:
\begin{equation}
\underline{\bm{\bm{\Sigma}}}_{\phi}=\left\langle \left\{ \delta\hat{\underline{a}}\delta\hat{\underline{a}}^{\dagger}\right\} _{\phi}\right\rangle \,,\label{eq:Correlations}
\end{equation}
where the anomalous correlations are: 
\begin{equation}
\left\langle \left\{ \delta\hat{a}_{\bm{k}}\delta\hat{a}_{\bm{k}'}\right\} _{\phi}\right\rangle =-\sum_{\bm{q}}\left[u_{\bm{k}\bm{q}}v_{\bm{k}'\bm{q}}^{*}+u_{\bm{k}'\bm{q}}v_{\bm{k}\bm{q}}^{*}\right]\left(\tilde{n}_{\bm{q}}+\frac{1}{2}\right),
\end{equation}
and the normal correlations are:
\begin{equation}
\left\langle \left\{ \delta\hat{a}_{\bm{k}}\delta\hat{a}_{\bm{k}'}^{\dagger}\right\} _{\phi}\right\rangle =\sum_{\bm{q}}\left[u_{\bm{k}\bm{q}}u_{\bm{k}'\bm{q}}^{*}+v_{\bm{k}'\bm{q}}v_{\bm{k}\bm{q}}^{*}\right]\left(\tilde{n}_{\bm{q}}+\frac{1}{2}\right)\,.
\end{equation}

The anomalous correlations like $\left\langle \left\{ \delta\hat{a}_{\bm{k}}\delta\hat{a}_{\bm{k}'}\right\} _{\phi}\right\rangle $
will average to zero after integrating over phase. However, these
are part of the initial quantum state, and since they contribute to
higher-order correlations as well, they must be included in computing
initial conditions for dynamical calculations. The non-phase-dependent
terms like $\left\langle \left\{ \delta\hat{a}_{\bm{k}}\delta\hat{a}_{\bm{k}'}^{\dagger}\right\} _{\phi}\right\rangle $
are the same for all phases. As an example, the average initial particle
number $\left\langle \hat{N}\right\rangle $, is an experimental observable.
From Eq (\ref{eq:NumberOperator}), since $\left\langle \hat{P}\right\rangle =0$,
\begin{equation}
N=N_{0}+\sum_{\bm{k}}\int d\bm{x}\left[\left|u_{\bm{k}}^{2}\left(\bm{x}\right)\right|\tilde{n}_{\bm{k}}+\left|v_{\bm{k}}^{2}\left(\bm{x}\right)\right|\left(\tilde{n}_{\bm{k}}+1\right)\right]\,\,.
\end{equation}
This allows one to solve implicitly for $N_{0}$ as a function of
the initial average particle number $N$, which typically involves
an iterative process, since the equations are inherently nonlinear.

Like the number itself, the number fluctuations are conserved and
depend on the quantum state preparation. One can calculate the number
fluctuations to second order in the field $\delta\hat{\Psi}$, which
gives from Eq (\ref{eq:NumberOperator}),
\begin{equation}
\delta N^{2}=\left\langle \left(\hat{N}-N\right)^{2}\right\rangle =\left\langle 2\hat{P}^{2}\right\rangle N_{0}\,.
\end{equation}
The initial state of the zeroth mode, $\hat{\rho}_{0}$, therefore
defines the number variance. If one chooses this to be the harmonic
oscillator ground state, with no excitations so that $\left\langle \hat{n}_{0}\right\rangle \equiv\tilde{n}_{0}=0$,
then $\left\langle \hat{P}^{2}\right\rangle =\left\langle \hat{Q}^{2}\right\rangle =1/2$
and the number fluctuations are just Poissonian, with $\delta N=\sqrt{N}$,
in the case that $N\approx N_{0}$. However, this is only one possible
initial condition, and it is independent of the states of the higher
lying modes.

The measured number fluctuations of a trapped BEC depend on the experiment.
With careful preparation, Poissonian fluctuations have been experimentally
obtained for $N\sim400$ \cite{bohi2009coherent}. Reductions to $50\%$
of the Poissonian level were achieved in small condensates of 100-200
atoms \cite{chuu2005direct}, increasing to Poissonian levels at around
$500$ atoms. Larger condensates are often super-Poissonian. Due to
noise in the evaporative cooling process and other issues, it is generally
difficult to reduce number fluctuations to $\pm\sqrt{N}$ for $N>1000$
with present technologies. 

These results may change in future. They depend on the details of
the state preparation in these isolated, trapped systems. To treat
the full range of cases quantitatively, one may choose the $\hat{b}_{0}$
condensate state to be in a squeezed or thermal state $\hat{\rho}_{0}$.
This leads to smaller or larger fluctuations respectively than $N_{0}$.
As a result, one has sub-Poissonian or super-Poissonian number statistics.
Any of these number fluctuation results are consistent with the techniques
described here.

\subsection{Homogeneous example}

As an example, the coefficients in the expansion for the homogeneous
case with $U=0$, confined in a volume $V$ are obtained. We assume
plane wave solutions with momentum $\hbar\bm{k}$, and periodic boundary
conditions. In this case, the mode functions can be written as $u_{\bm{k}}(\bm{x})=u_{\bm{k}}\psi_{\bm{k}}\left(\bm{x}\right)$,
$v_{\bm{k}}(\bm{x})=v_{\bm{k}}\psi_{\bm{k}}\left(\bm{x}\right)$.
Here $\bm{k}$ is a vector in one or more dimensions, we define :
$\psi_{\bm{k}}\left(\bm{x}\right)=\exp(i\bm{k}\cdot\bm{x})/\sqrt{V}\,,$and
we can write that: 
\begin{equation}
\delta\hat{\Psi}\left(\bm{x}\right)=\frac{1}{\sqrt{V}}\sum_{\bm{k}}\left[\hat{b}_{\bm{k}}u_{\bm{k}}e^{i\bm{k}\cdot\bm{x}}-\hat{b}_{\bm{k}}^{\dagger}v_{\bm{k}}^{*}e^{-i\bm{k}\cdot\bm{x}}\right]\,.
\end{equation}

This leads to the simple result that $\hat{a}_{\bm{k}}=\alpha_{\bm{k}}^{(0)}+\hat{b}_{\bm{k}}u_{\bm{k}}-\hat{b}_{-\bm{k}}^{\dagger}v_{-\bm{k}}^{*}$,
after expanding $\delta\hat{\Psi}\left(\bm{x}\right)$. Then, solving
Eq (\ref{eq:BdG}) by introducing the kinetic energy, $E_{\bm{k}}$
and phonon energy $\epsilon_{\bm{k}}$, defined in frequency units,
one obtains well-known results, which are repeated here for completeness:
\begin{eqnarray}
E_{\bm{k}} & = & \frac{\hbar^{2}\left|\bm{k}\right|^{2}}{2m}\nonumber \\
\epsilon_{\bm{k}} & = & \sqrt{E_{\bm{k}}\left(E_{\bm{k}}+2gn_{0}\right)}\,.
\end{eqnarray}
The corresponding coefficients $u_{\bm{k}}$, $v_{\bm{k}}$ are defined
such that, for $\bm{k}\neq0$:
\begin{eqnarray}
u_{\bm{k}} & = & \frac{\epsilon_{\bm{k}}+E_{\bm{k}}}{2\sqrt{\epsilon_{\bm{k}}E_{\bm{k}}}}\nonumber \\
v_{\bm{k}} & = & \frac{\epsilon_{\bm{k}}-E_{\bm{k}}}{2\sqrt{\epsilon_{\bm{k}}E_{\bm{k}}}}\,.
\end{eqnarray}

For $\bm{k}\neq0$, there are no singularities, and we must have 
\begin{equation}
\left|u_{\bm{k}}\right|^{2}-\left|v_{\bm{k}}\right|^{2}=1\,.\label{eq:u2_v2_relation}
\end{equation}
The following algebraic relations also hold: 
\begin{eqnarray}
\left|u_{\bm{k}}\right|^{2}+\left|v_{\bm{k}}\right|^{2} & = & \frac{E_{\bm{k}}+gn_{0}}{\epsilon_{\bm{k}}}\nonumber \\
u_{\bm{k}}v_{\bm{k}} & = & \frac{gn_{0}}{2\epsilon_{\bm{k}}}\,.
\end{eqnarray}

The result for $\bm{k}=0$ is different, as explained in the previous
subsection. If one substitutes into the above equation, which are
only valid for $\bm{k}\neq0$, the expansion coefficients would diverge.
Instead, using the zeroth mode method given above, one finds that:
\begin{equation}
\psi_{0}=\Phi_{0}=\frac{1}{\sqrt{V}}\,.
\end{equation}
With this choice, the corresponding coefficient is $\alpha=gn_{0}$,
with coefficients $u_{0}=1$, $v_{0}=0$. Similarly, from Eq (\ref{eq:nonlinearChemPotSoln}),
$\mu_{2}=\alpha/N_{0}=g/V$.

\subsection{Variances and quantum squeezing}

In the simplest case of a homogeneous BEC, the relationship in Eq
(\ref{eq:u2_v2_relation}) shows that the mode coefficients for a
single mode can be written:

\begin{eqnarray}
u_{\bm{k}} & = & cosh(r_{\bm{k}})\nonumber \\
v_{\bm{k}} & = & sinh(r_{\bm{k}})\,.
\end{eqnarray}
Next, since the squeezing transformation is defined by $\delta\hat{a}_{\bm{k}}=u_{\bm{k}}\hat{b}_{\bm{k}}-v_{\bm{k}}^{*}\hat{b}_{-\bm{k}}^{\dagger}\,,$
for $\bm{k}\neq0$ we can introduce new odd and even operators $\hat{a}_{\bm{k}\pm}=\left[\delta\hat{a}_{\bm{k}}\pm\delta\hat{a}_{-\bm{k}}\right]/\sqrt{2}$
and $\hat{b}_{\bm{k}\pm}=\left[\hat{b}_{\bm{k}}\pm\hat{b}_{-\bm{k}}\right]/\sqrt{2}$
such that 
\begin{equation}
\hat{a}_{\bm{k}\pm}=u_{\bm{k}}\hat{b}_{\bm{k}\pm}\mp v_{\bm{k}}^{*}\hat{b}_{\bm{k}\pm}^{\dagger}\,.
\end{equation}
This is an example of a single-mode quantum squeezing transformation,
which is a Gaussian state. For $\bm{k}=0$ we define $\hat{a}_{0+}=\hat{a}_{0}$,
and there is no odd mode. The quantum statistics are given by the
quadrature operators of the odd and even modes, 
\begin{eqnarray}
\hat{P}_{\bm{k}\pm} & = & \left(\hat{a}_{\bm{k}\pm}+\hat{a}_{\bm{k}\pm}^{\dagger}\right)/\sqrt{2}\nonumber \\
\hat{Q}_{\bm{k}\pm} & = & i\left(\hat{a}_{\bm{k}\pm}-\hat{a}_{\bm{k}\pm}^{\dagger}\right)/\sqrt{2}.
\end{eqnarray}

In this squeezed state, the two symmetrically ordered quadrature variances
are given by:

\begin{eqnarray}
\left\langle \hat{P}_{\bm{k}\pm}^{2}\right\rangle  & = & \left[\tilde{n}_{\bm{k}}+\frac{1}{2}\right]e^{\mp2r_{\bm{k}}}\nonumber \\
\left\langle \hat{Q}_{\bm{k}\pm}^{2}\right\rangle  & = & \left[\tilde{n}_{\bm{k}}+\frac{1}{2}\right]e^{\pm2r_{\bm{k}}}\,.
\end{eqnarray}
The zeroth mode term in the second order Hamiltonian is normally responsible
for the large phase fluctuations in Bogoliubov theory. Since these
generate quadratic terms in $\hat{P}$, they are canceled by the second
order chemical potential. This is similar to the fact that linear
terms in $\hat{P}$ are canceled by a choice of linear chemical potential.

Combining the second order zero-momentum terms together, we see that
the zeroth mode contribution to the Hamiltonian is 
\begin{equation}
\hat{H}_{0}^{(2)}=N_{0}\left[\frac{g}{V}-\mu_{2}\right]\hat{P}^{2}.
\end{equation}
This means that we should choose that $\mu_{2}=g/V$, as also shown
in the previous section, so that $\mu_{1}=0$ in this case. This leaves
no residual term in the quadratic Hamiltonian that depends on $\hat{P}$,
thus removing all divergences. Since the zeroth mode has zero energy,
the mode energy is unchanged, and is still given by: 
\begin{equation}
\hat{H}^{(2)}=\sum_{\bm{k}\neq0}\epsilon_{\bm{k}}\hat{b}_{\bm{k}}^{\dagger}\hat{b}_{\bm{k}}\,.
\end{equation}

\section{States and phase-space representations }

The results given above describe the small quantum and thermal fluctuations
of a low temperature Bose condensate. More generally, large dynamical
changes can occur during the evolution that occurs after a change
in the Hamiltonian, for example, after a beam-splitter, or a quench
in which the coupling is changed. Under these circumatsances, it is
no longer possible to only treat a quadratic approximate Hamiltonian,
and more suitable methods must be used. Of these, some of the most
useful are phase-space representations, which are applicable even
to very large Hilbert spaces of three-dimensional condensates, either
with \cite{egorov2011long,opanchuk2013functional} or without \cite{norrie_ballagh_05,deuar2007correlations}
dissipation.

Wigner\cite{wigner_32} developed the first quantum phase-space representation.
Subsequently, Moyal\cite{moyal_49} showed how to use phase-space
mappings to calculate quantum dynamics. Variations of these techniques
were obtained by Husimi\cite{husimi_40}, Glauber\cite{glauber_63},
Sudarshan\cite{sudarshan_63}, Lax\cite{lax_louisell_67}, and others,
by using different operator orderings. These are generically called
quasi-probability distributions or representations, although these
early techniques have no stochastic process that corresponds directly
to the quantum dynamics, unless they are truncated or approximated.

Other methods exist that do have a stochastic process, but require
more sophisticated techniques, such as the positive-P representation
\cite{DrummondGardinerP1980}. Here we will show how to utilize the
results obtained above to initialize either the Wigner or the positive-P
phase-space method.

\subsection{State transformations}

The previous section diagonalized the grand canonical Hamiltonian
approximately, using a displacement followed by a squeezing transformation
on the operators. This is a unitary operator transformation which
depends on the condensate field, and is written as: 
\begin{equation}
\hat{b}_{\bm{k}}=\hat{U}_{\phi}\hat{a}_{\bm{k}}\hat{U}_{\phi}^{\dagger}\,,
\end{equation}
where $\hat{U}_{\phi}=\hat{U}\left[\Psi_{\phi}\right]=\hat{U}_{S}\hat{D}$.
It has the property that it transforms the grand canonical Hamiltonian
to a diagonal form within the Bogoliubov approximation, i.e, 
\begin{equation}
\hat{K}^{(2)}=\sum_{\bm{k}}\epsilon_{\bm{k}}\hat{b}_{\bm{k}}^{\dagger}\hat{b}_{\bm{k}}=\hat{U}_{\phi}\sum_{\bm{k}}\epsilon_{\bm{k}}\hat{a}_{\bm{k}}^{\dagger}\hat{a}_{\bm{k}}\hat{U}_{\phi}^{\dagger}\,.
\end{equation}
As a result, if $\left|0\right\rangle $ is the original bosonic vacuum
state, then the approximate interacting ground state $\left|g_{\phi}\right\rangle $
is: 
\begin{equation}
\left|g_{\phi}\right\rangle =\hat{U}_{\phi}\left|0\right\rangle \,.
\end{equation}
Combining these results together, if $\hat{\rho}_{th}$ is a thermal
density matrix for free phonons with $\bm{k}\neq0$, and $\hat{\rho}_{0}$
is the density matrix of the condensate mode, then the initial state
of the BEC is a phase-integral: 
\begin{equation}
\hat{\rho}=\frac{1}{2\pi}\int d\phi\hat{U}_{\phi}\hat{\rho}_{th}\hat{\rho}_{0}\hat{U}_{\phi}^{\dagger}\,.
\end{equation}

This is a bosonic Gaussian state for every condensate phase, and hence
can be easily represented using phase-space techniques. The initial
state can then be evolved in time when the Hamiltonian is changed,
to model quantum dynamical experiments.

\subsection{Wigner representation}

Of these classical phase-space mappings a truncated version of the
Wigner distribution is generally useful in quantum field dynamics,
and it is often used for describing quantum optical and BEC systems\cite{drummond_hardman_93,steel_olsen_98,blakie_bradley_08,martin2010quantum,corney2006many,egorov2011long}.
This employs a $1/N$ expansion, truncated after second derivative
terms are obtained, which limits its applicability to large condensates.
In the case of Bose-Einstein condensates, this method was used for
the treatment of a BEC in an optical lattice \cite{Ruostekowski2005,Isella2005},
with Bogoliubov initial conditions. This treatment was in one dimension
and used a different approach to describing the condensate mode fluctuations.

The Wigner distribution is a functional distribution which is a symmetrically
ordered representation of the quantum field \cite{DrummondHardman1993,Steel1998},
and in this case it is a mixture of terms at each phase $\phi$: 
\begin{equation}
W_{\phi}\left[\Psi\right]=\frac{1}{2\pi}\int d\phi W_{\phi}\left[\Psi\right].
\end{equation}
For any symmetrically ordered moment of field operators, $f\left(\hat{\Psi},\hat{\Psi}^{\dagger}\right)$,
one has: 
\begin{equation}
\left\langle \left\{ f\left(\hat{\Psi},\hat{\Psi}^{\dagger}\right)\right\} \right\rangle =\frac{1}{2\pi}\int d\phi\int\mathcal{D}\Psi f\left(\Psi,\Psi^{*}\right)W_{\phi}\left[\Psi\right]\,.
\end{equation}
Here $\left\{ \right\} $ is used to indicate symmetric ordering,
and $\mathcal{D}\Psi$ is a functional integration measure. Since
the Bogoliubov operators are simply a linear combination of field
mode operators, the Wigner representation can be used for either phonons
or bosons. We note here that while a positive Wigner distribution
does not always exist, $W_{\phi}\left[\Psi\right]$ does exist for
all Gaussian states, and the initial vacuum, squeezed or thermal state
are all examples of a Gaussian state.

If thermal phonons are described by the phonon operators $\hat{b}_{\bm{k}}$,
they have a Wigner representation correspondence such that 
\begin{equation}
\hat{b}_{\bm{k}}\sim\beta_{\bm{k}}\,.
\end{equation}
We define $\tilde{n}_{\bm{k}}$ as the thermal occupation number of
phonon modes at temperature T and energy $\epsilon_{\bm{k}}$, for
$\bm{k}\neq0$, and a vacuum state for $\bm{k}=0$ in the case of
Poissonian number fluctuations. Of course, other choices are needed
for the condensate fluctuations The Wigner representation of the approximate
initial state is a Gaussian, positive probability distribution, with
a complex, symmetrically ordered variance in each phonon mode given
by: 
\begin{equation}
\left\langle \left|\beta_{\bm{k}}\right|^{2}\right\rangle =\tilde{n}_{\bm{k}}+\frac{1}{2}\,.
\end{equation}

The Wigner initial conditions are therefore thermal and quantum stochastic
fluctuations of the phonons, and the initial condition is simply a
sum over Wigner squeezed state amplitudes, so for $\phi=0$: 
\begin{equation}
\Psi_{W}=\Psi_{0}\left(\bm{x}\right)+\sum\left(u_{\bm{k}}\left(\bm{x}\right)\beta_{\bm{k}}-v_{\bm{k}}^{*}\left(\bm{x}\right)\beta_{\bm{k}}^{*}\right)\,.
\end{equation}

Here, the initial Wigner distribution in this case is a Gaussian state,
because the changed fluctuations from the usual vacuum state correspond
to quantum squeezing. It is well-known that one and two-mode squeezed
states all have a positive Wigner representation, defined completely
by their second-order statistics.

For a general condensate, the result for the initial mode amplitudes
is that 
\begin{equation}
\alpha_{\bm{k}}=\alpha_{\bm{k}}^{(0)}+u_{\bm{k}\bm{q}}\beta_{\bm{q}}-v_{\bm{k}\bm{q}}^{*}\beta_{\bm{q}}^{*}\,.
\end{equation}

In the homogeneous case, these results provide a rigorous justification
for a previously suggested procedure \cite{Ruostekowski2005,isella2005nonadiabatic},
while more generally they lead to a new treatment of the condensate
mode statistics using the correct conjugate mode functions. In particular,
for a non-uniform condensate one generally has $v_{0}\neq0$, in order
to satisfy the relevant eigenvalue equation, Eq (\ref{eq:zero-mode}).
As a result, even if the condensate is in the Bogoliubov ground state,
the corresponding quantum fluctuations are squeezed. They will in
general differ from the quantum fluctuations of a coherent state of
the same mode.

Whether this is an important issue or not depends on the coupling
and the trap. In some three-dimensional inhomogeneous trap BEC experiments
it is possible to even assume that $v_{k}\approx0$, since most of
the phonon excitations are in a thermal cloud external to the condensate,
where they behave to a good approximation as a noninteracting Bose
gas \cite{holzmann1999precision,gerbier2004experimental}.

In the resulting truncated Wigner quantum dynamical calculations,
we do not linearize, since there can be very large changes from the
equilibrium state. The chemical potential terms conserve the total
particle number, and have no effect on the dynamical evolution of
any number-conserving observable. Accordingly, it is irrelevant whether
they are included or omitted. The Wigner dynamical equations are very
similar to those obtained in the Heisenberg picture from the original
Hamiltonian. Defining couplings in frequency units as $\tilde{g}\equiv g/\hbar$,
$\tilde{U}\equiv U/\hbar$, then 
\begin{equation}
\partial_{t}\Psi_{W}=\frac{i\hbar}{2m}\nabla^{2}\Psi_{W}-i\left[\tilde{g}\left|\Psi_{W}\right|^{2}+\tilde{U}\right]\Psi_{W}\,.
\end{equation}

There are also cut-off-dependent frequency shifts, but as these simply
behave as chemical potential terms, they have no effect on observables,
leaving an equation in the form of a Gross-Pitaevskii equation with
initial quantum fluctuations. If there is additional gain or loss
as well, then additional noise terms are required to preserve commutation
relations \cite{opanchuk2013functional}. Methods of numerical integration
using Fourier transforms or other spectral methods and projections
are described in the literature \cite{blakie2008numerical}.

\subsubsection{Homogeneous example}

In the homogeneous case, this simply gives the result that 
\begin{equation}
\alpha_{\bm{k}}=\alpha_{\bm{k}}^{(0)}+u_{\bm{k}}\beta_{\bm{k}}-v_{\bm{k}}\beta_{-\bm{k}}^{*}\,.
\end{equation}

These Wigner amplitudes correspond to two-mode squeezed phonon states
\cite{Yuen1976}, with correlations between the opposite momenta.
We can return to the original mode variables to construct the initial
conditions for the equations of motion, so that the field expansion
is just a Fourier transform: 
\begin{equation}
\Psi_{W}=\frac{1}{\sqrt{V}}\sum_{\bm{k}}\alpha_{\bm{k}}e^{i\bm{k}\bm{x}}.
\end{equation}

\subsection{Positive-P representation}

The positive $P$-representation is a useful alternative to the Wigner
approach. This has a stochastic process with no truncation. It is
widely used for simulating interacting quantum fields in quantum optics
\cite{carter1987squeezing} and for transients in BECs \cite{deuar2007correlations,drummond1999quantum}.
In the absence of damping the time-scales it is useful for are limited
owing to the growth of sampling error, but it can provide an accurate
quantum simulation of initial transients.

In this representation, a single quantum mode has two complex amplitudes,
which we can label $\boldsymbol{\alpha}$ and $\boldsymbol{\alpha}^{+}$.
Here $\boldsymbol{\alpha}$ corresponds to an annihilation operator,
while $\boldsymbol{\alpha}^{+}$ corresponding to a creation operator.
On average, $\boldsymbol{\alpha}^{+}$ is complex conjugate to $\boldsymbol{\alpha}$.
The expansion of the density matrix is then: 
\begin{equation}
\hat{\rho}=\frac{1}{2\pi}\int d\phi\int\int P_{\phi}(\boldsymbol{\alpha},\boldsymbol{\alpha}^{+})\,\hat{\Lambda}(\boldsymbol{\alpha},\boldsymbol{\alpha}^{+})d^{2M}\boldsymbol{\alpha}\,d^{2M}\boldsymbol{\boldsymbol{\alpha}^{+}}\,.\label{eq:drummond_+P}
\end{equation}
where $\hat{\Lambda}(\boldsymbol{\alpha},\boldsymbol{\alpha}^{+})=\left|\boldsymbol{\alpha}\right\rangle \left\langle \boldsymbol{\alpha}^{+*}\right|/\left\langle \boldsymbol{\alpha}^{+*}\right|\left.\boldsymbol{\alpha}\right\rangle $
is a coherent state projector, and $\left|\boldsymbol{\alpha}\right\rangle $
is an eigenstate of $\hat{\bm{a}}$.

The construction of the Gaussian states of the BEC in the positive-P
representation can be carried out in several ways, as the representation
is not unique. One method is to base the representation on a canonical
form based on the Q-function \cite{DrummondGardinerP1980} that exists
for any density matrix, and therefore can be used to generate squeezed
states \cite{olsen2009numerical}. Here we use a more compact and
efficient form available in the case of Gaussian states \cite{kiesewetter2017pulsed},
by use of the symmetrically ordered correlations defined in Eq (\ref{eq:Correlations})
of the previous section.

The corresponding normally ordered correlations can be calculated
from the symmetrically ordered correlations, so that for each phase
angle $\underline{\bm{\bm{\Sigma}}_{N}}=\underline{\bm{\bm{\Sigma}}_{\phi}}-\underline{I}/2$,
where $\underline{\bm{\bm{\Sigma}}_{\phi}}$ is given in Eq ((\ref{eq:Correlations}).
This can be factorized as a matrix square root that may have complex
values, with: 
\begin{equation}
\underline{\bm{\bm{\Sigma}}_{N}}=\underline{\bm{\bm{\sigma}}_{P}}\underline{\bm{\bm{\sigma}}_{P}}^{T}.
\end{equation}
The corresponding positive-P variables are then initialized according
to the prescription that: 
\begin{equation}
\left[\begin{array}{c}
\bm{\alpha}\\
\bm{\alpha^{+}}
\end{array}\right]=\left[\begin{array}{c}
\bm{\alpha}_{0}\\
\bm{\alpha}_{0}^{*}
\end{array}\right]+\underline{\bm{\bm{\sigma}}_{P}}\bm{\zeta}.
\end{equation}

The resulting positive-P equations for interacting Bose-Einstein condensates
(BECs) have a relatively simple form. We transform the initial coherent
amplitudes from mode to position space, giving rise to an equivalent
set of c-number field amplitudes 
\begin{equation}
\Psi=\sum_{\bm{k}}\alpha_{\bm{k}}u_{\bm{k}}^{(0)}\left(\bm{x}\right),
\end{equation}
and 
\begin{equation}
\Psi^{+}=\sum_{\bm{k}}\alpha_{\bm{k}}^{+}u_{\bm{k}}^{(0)*}\left(\bm{x}\right)\,.
\end{equation}
After reconstructing the corresponding quantum fields, $\Psi$and
$\Psi^{\dagger}$, these obey two coupled stochastic partial differential
equations: 
\begin{equation}
\partial_{t}\Psi=i\left[\frac{\hbar}{2m}\nabla^{2}-\tilde{g}\Psi^{+}\Psi-\tilde{U}+\sqrt{i\tilde{g}}\xi\right]\Psi,
\end{equation}

\begin{equation}
\partial_{t}\Psi^{+}=-i\left[\frac{\hbar}{2m}\nabla^{2}-\tilde{g}\Psi^{+}\Psi-\tilde{U}+\sqrt{-i\tilde{g}}\xi^{+}\right]\Psi^{+}.
\end{equation}
Here $\xi$ and $\xi^{+}$ are real Gaussian noises, independent at
each time step (of length $\Delta t$) and lattice point, with standard
deviations $1/\sqrt{\Delta V\Delta t}$. These equations are similar
to the mean-field Gross-Pitaevskii equations in a doubled phase-space,
with the addition of independent noise. This incorporates all quantum
effects, provided boundary conditions are met at large phase-space
radius \cite{deuar2006first}.

Simulations of this type do not require assumptions about which modes
are occupied and which are not. All quantum modes up to a cut-off
$\bm{k}_{c}$ that depends on the lattice, are included. With this
approach, short time-scale transient dynamics are accurately computable
from first principles, even for large interacting BECs \cite{deuar2007correlations}.
These are stochastic partial differential equations, and both algorithms
\cite{Werner:1997} and software \cite{Kiesewetter2016} for these
equations are are available.

\subsubsection{Homogeneous example}

To generate this in the positive-P distribution, we note that due
to normal ordering, the variances are zero in a coherent state. For
squeezed states the variances are: 
\begin{equation}
\Sigma_{\bm{k}\pm,p}=\left\langle :\hat{P}_{\bm{k}\pm}^{2}:\right\rangle =\left[\frac{1}{2}+\tilde{n}_{\bm{k}}\right]e^{\mp2r_{\bm{k}}}-\frac{1}{2}
\end{equation}
and
\begin{equation}
\Sigma_{\bm{k}\pm,q}=\left\langle :\hat{Q}_{\bm{k}\pm}^{2}:\right\rangle =\left[\frac{1}{2}+\tilde{n}_{\bm{k}}\right]e^{\pm2r_{\bm{k}}}-\frac{1}{2}.
\end{equation}

The phase-space standard deviations are defined here as $\sigma_{\bm{k}\pm,p}=\sqrt{\Sigma_{\bm{k}\pm,p}}$
, and $\sigma_{\bm{k}\pm,q}=\sqrt{\Sigma_{\bm{k}\pm,q}}$. Since either
normally-ordered variance could be negative, the square roots can
be imaginary. Hence these are generated by two noise fields that may
not be complex conjugate, where: 
\begin{equation}
\alpha_{\bm{k}\pm}=\frac{1}{\sqrt{2}}\left[\zeta_{\bm{k}\pm,p}\sigma_{\bm{k}\pm,p}-i\zeta_{\bm{k}\pm,q}\sigma_{\bm{k}\pm,q}\right]
\end{equation}

\begin{equation}
\alpha_{\bm{k}\pm}^{+}=\frac{1}{\sqrt{2}}\left[\zeta_{\bm{k}\pm,p}\sigma_{\bm{k}\pm,p}+i\zeta_{\bm{k}\pm,q}\sigma_{\bm{k}\pm,q}\right]
\end{equation}
and $\zeta_{\bm{k}\pm,p(q)}$ are uncorrelated, real Gaussian noise
terms of unit variance. These should be recombined to give the mode
amplitudes, via: 
\begin{equation}
\alpha_{\bm{k}}=\alpha_{\bm{k}}^{(0)}+\frac{1}{\sqrt{2}}\left[\alpha_{\bm{k}+}+\alpha_{\bm{k}-}\right],
\end{equation}

\begin{equation}
\alpha_{\bm{k}}^{+}=\alpha_{\bm{k}}^{(0)*}+\frac{1}{\sqrt{2}}\left[\alpha_{\bm{k}+}^{+}+\alpha_{\bm{k}-}^{+}\right]\,.
\end{equation}

The occupation number in a squeezed thermal state is therefore 
\begin{equation}
\left\langle \hat{n}_{\bm{k}}\right\rangle =\left[\frac{1}{2}+\tilde{n}_{\bm{k}}\right]cosh(r_{\bm{k}})-\frac{1}{2}\,.
\end{equation}

\section{Conclusion}

We have introduced the concept of a nonlinear chemical potential for
the thermal equilibrium state of a Bose-Einstein condensate. The resulting
grand canonical Hamiltonian is shown to be locally bounded below for
small fluctuations. The bound is proved by using a Bogoliubov transformation
without divergences, including the condensate mode term. This allows
a complete unitary transformation on the field, which is essential
in treating finite condensates that are not in the large volume, thermodynamic
limit. The result is a diagonal phonon Hamiltonian, including the
condensate mode. This is important, in view of theorems that require
theories to be gapless and conserving \cite{goldstone1961field,hugenholtz1959ground}.

Our approach is then used to calculate the equilibrium density matrix,
which can be used for subsequent nonperturbative dynamical calculations.
The density matrix is phase-independent, and consists of a probabilistic
mixture of Gaussian states. Such quantum states have a distribution
over particle number. The number variance can be chosen to correspond
to the initial quantum ensemble. The advantage of the technique over
methods that use number state initial conditions, is that a grand
canonical ensemble is more typical of BEC experiments. This approach
also resolves a problem with using older Bogoliubov methods in which
the condensate mode is omitted, since it allows the quantum state
of the condensate mode to be included in a consistent way.

As an example, we use the initial thermal quantum density matrix to
calculate initial conditions for stochastic phase-space simulation
of the quantum dynamics of a BEC. Both the Wigner and positive-P phase-space
distribution methods are described. Quantum and thermal fluctuations
are included. We expect the nonlinear chemical potential approach
to be of general applicability in cases of broken symmetry, and it
is especially suitable for the calculation of the quantum dynamics
of correlations and entanglement in mesoscopic BEC experiments. Detailed
quantum dynamical simulations using these methods and their agreement
with experimental BEC interferometry will be reported elsewhere \cite{Opanchuk2018simulation}.

\ack{}{}

PDD thanks the hospitality of the Institute for Atomic and Molecular
Physics (ITAMP) at Harvard University, supported by the NSF. 

\bibliographystyle{iopart-num}
\bibliography{Divergenceless,drummond}

\end{document}